\documentclass[12pt]{article}
\usepackage{graphicx,rotating,subfigure,amsmath}
\newcommand{\BABARPubYear}     {07}
\newcommand{\BABARPubNumber}  {128}
\newcommand{\SLACPubNumber} {13011}
\usepackage{relsize}
\RequirePackage{xspace}
\def\babar{\mbox{\slshape B\kern-0.1em{\smaller A}\kern-0.1em
    B\kern-0.1em{\smaller A\kern-0.2em R}}}
\def\CP    {\ensuremath{C\!P}\xspace}

\def\epem  {\ensuremath{e^+e^-}\xspace}
\def\q     {\ensuremath{q}\xspace}

\def\qqbar {\ensuremath{q\overline q}\xspace}
\def\u     {\ensuremath{u}\xspace}

\def\d     {\ensuremath{d}\xspace}

\def\s     {\ensuremath{s}\xspace}

\def\b     {\ensuremath{b}\xspace}

\def\piz   {\ensuremath{\pi^0}\xspace}
\def\pip   {\ensuremath{\pi^+}\xspace}
\def\pim   {\ensuremath{\pi^-}\xspace}

\def\Kbar  {\kern 0.2em\overline{\kern -0.2em K}{}\xspace}

\def\Kz    {\ensuremath{K^0}\xspace}
\def\Kzb   {\ensuremath{\Kbar^0}\xspace}
\def\KzKzb {\ensuremath{\Kz \kern -0.16em \Kzb}\xspace}
\def\Kp    {\ensuremath{K^+}\xspace}
\def\Km    {\ensuremath{K^-}\xspace}
\def\KpKm  {\ensuremath{\Kp \kern -0.16em \Km}\xspace}
\def\KS    {\ensuremath{K^0_{\scriptscriptstyle S}}\xspace} 
\def\Kstarz  {\ensuremath{K^{*0}}\xspace}
\def\Kstarzb {\ensuremath{\Kbar^{*0}}\xspace}

\def\B       {\ensuremath{B}\xspace}
\def\Bbar    {\kern 0.18em\overline{\kern -0.18em B}{}\xspace}
\def\Bb      {\ensuremath{\Bbar}\xspace}
\def\BB      {\ensuremath{B\Bbar}\xspace} 
\def\Bzb     {\ensuremath{\Bbar^0}\xspace}

\mathchardef\Upsilon="7107
\def\Y#1S{\ensuremath{\Upsilon{(#1S)}}\xspace}
\def\FourS {\Y4S}
\def\mes        {\mbox{$m_{\rm ES}$}\xspace}
\def\DeltaE     {\mbox{$\Delta E$}\xspace}
\newcommand{\gev}{\ensuremath{\mathrm{\,Ge\kern -0.1em V}}\xspace}
\newcommand{\mev}{\ensuremath{\mathrm{\,Me\kern -0.1em V}}\xspace}
\newcommand{\gevc}{\ensuremath{{\mathrm{\,Ge\kern -0.1em V\!/}c}}\xspace}
\newcommand{\mevc}{\ensuremath{{\mathrm{\,Me\kern -0.1em V\!/}c}}\xspace}
\newcommand{\gevcc}{\ensuremath{{\mathrm{\,Ge\kern -0.1em V\!/}c^2}}\xspace}
\newcommand{\mevcc}{\ensuremath{{\mathrm{\,Me\kern -0.1em V\!/}c^2}}\xspace}

\def\to                 {\ensuremath{\rightarrow}\xspace}
\newcommand{\stat}{\ensuremath{\mathrm{(stat)}}\xspace}
\newcommand{\syst}{\ensuremath{\mathrm{(syst)}}\xspace}
\def\pep2{PEP-II}
\begin{document}
\begin{flushright}
SLAC-PUB-\SLACPubNumber \\
\babar-TALK-\BABARPubYear/\BABARPubNumber\\
November 2007
\end{flushright}
\par\vskip 4cm
\begin{center}
\Large \bf Charmless Hadronic {\boldmath $\B$} Decays at \babar $^*$
\end{center}
\bigskip
\begin{center}
Gagan B. Mohanty\\
{\em Representing the \babar\ Collaboration}\\
Department of Physics, University of Warwick, Coventry CV4 7AL, UK
\end{center}
\bigskip \bigskip
\begin{abstract}
We report recent measurements of branching fractions and charge asymmetries
of charmless hadronic \B\ decays using the data collected with the \babar\
detector at the \pep2\ asymmetric energy \epem\ collider.
\end{abstract}
\vfill
\begin{center}
Presented at XII International Conference on Hadron Spectroscopy (HADRON07),\\
Frascati, Italy, 8-13 October 2007.\\
Submitted to Frascati Physics Series (Proceedings)
\end{center}
\vspace{1.0cm}
\begin{center}
{\em Stanford Linear Accelerator Center, Stanford University, 
Stanford, CA 94309} \\ \vspace{0.1cm}\hrule\vspace{0.1cm}
$^*$Work supported in part by Department of Energy contract DE-AC02-76SF00515.
\end{center}
\newpage
\section{Introduction
}
\B\ meson decays to hadronic final states without a charm
quark are an important probe of the Standard Model (SM) of particle
physics. The so-called charmless decays play a key role in testing the
Cabibbo-Kobayashi-Maskawa (CKM) predictions of charge-parity (\CP)
violation, with sensitivity to the three angles $\alpha$, $\beta$ and
$\gamma$ of the Unitarity Triangle. The dominant contributors to this
class of \B\ decays are ``penguin'' decays mediated by $\b\to\s$ and
$\b\to\d$ process involving a virtual loop with the emission of a gluon
and the CKM-suppressed $\b\to\u$ tree diagram. Due to the presence of
an extra strange quark, the first diagram contributes only to final
states containing an odd number of kaons; while the latter two result
in final states with no, or an even number of, kaons. Penguin decays
provide an ideal environment to look for new physics (NP) with possible
contributions from non-SM particles in the loop, while the interference
between tree and penguin amplitudes (of comparable magnitudes) allows
to search for direct \CP\ violation~\cite{Kpi-paper}. Furthermore,
studies of these decay processes can be used to constrain varieties
of theoretical models of \B\ decays based on factorization,
perturbative QCD, and SU(3) flavor symmetry.

In these proceedings, we summarize most recent results on charmless hadronic
\B\ decays~\cite{charge} culminating in three-body, quasi-two-body
(Q2B), or other multibody final states; studied using \epem\ collision data
collected with the \babar\ detector~\cite{babar} near the \FourS\ resonance.
The results should be considered preliminary, unless a journal reference
is given.
\section{Analysis Method
}
The challenge in studying charmless hadronic \B\ decays is to extract
a small signal (typical branching fraction is of the order of $10^{-6}$)
out of a sea of background events. The continuum light-quark production,
$\epem\to\qqbar\,(\q=u,d,s,c)$, forms the most dominant background
component. It is suppressed by exploiting the difference in event
topology - \B\ mesons are produced almost at rest resulting in a
spherical event, while the light-quark pairs tend to have a
jetlike shape owing to the large available kinetic energy - and by
utilizing flavor and decay-time information of \B\ meson candidates.
Particle identification plays a crucial role in separating
charged pions from kaon track candidates. This becomes particularly
important against the background emanating from \B\ decays with
similar hadronic final states. Backgrounds from final states with
charm quarks are suppressed by invariant-mass vetoes on charmonia
and $D$ mesons. The signal yield is extracted by performing an
unbinned maximum-likelihood (ML) fit to event-shape variables (usually
combined to a Fisher discriminant ${\cal F}$ or an artificial Neural
Network NN), and kinematic quantities that make use of precise
beam-energy information and energy-momentum conservation. The
kinematic variables are the difference \DeltaE\ between the energy
of the reconstructed \B\ candidate and the beam energy ($E_{\rm beam}$),
and the beam-energy substituted mass $\mes\equiv\sqrt{E^2_{\rm beam}-
{\bf p}^2_\B}$, where ${\bf p}_\B$ is the momentum of the \B\ candidate
[here all quantities are calculated in the \FourS\ rest frame]. Where
available, the invariant-mass and angular variables of Q2B resonances
are used to further enhance background suppression. For signal modes
with a significant yield, the \CP\ violation or charge asymmetry is
measured using $A_{\CP}=\frac{N_{\Bb}-N_\B}{N_{\Bb}+N_\B}$, where
$N_{\Bb}$ and $N_\B$ correspond to the number of $\Bb$\, (\Bzb\ or
$B^-$) and $B$\, ($\B^0$ or $B^+$) decays detected in the inclusive
yield, respectively.
\section{Experimental Results
}
\subsection{Three-body Decay $B^+\to K^+K^-\pi^+$
}
\begin{figure}[!htb]
    \begin{center}
        {\includegraphics[width=.48\textwidth]{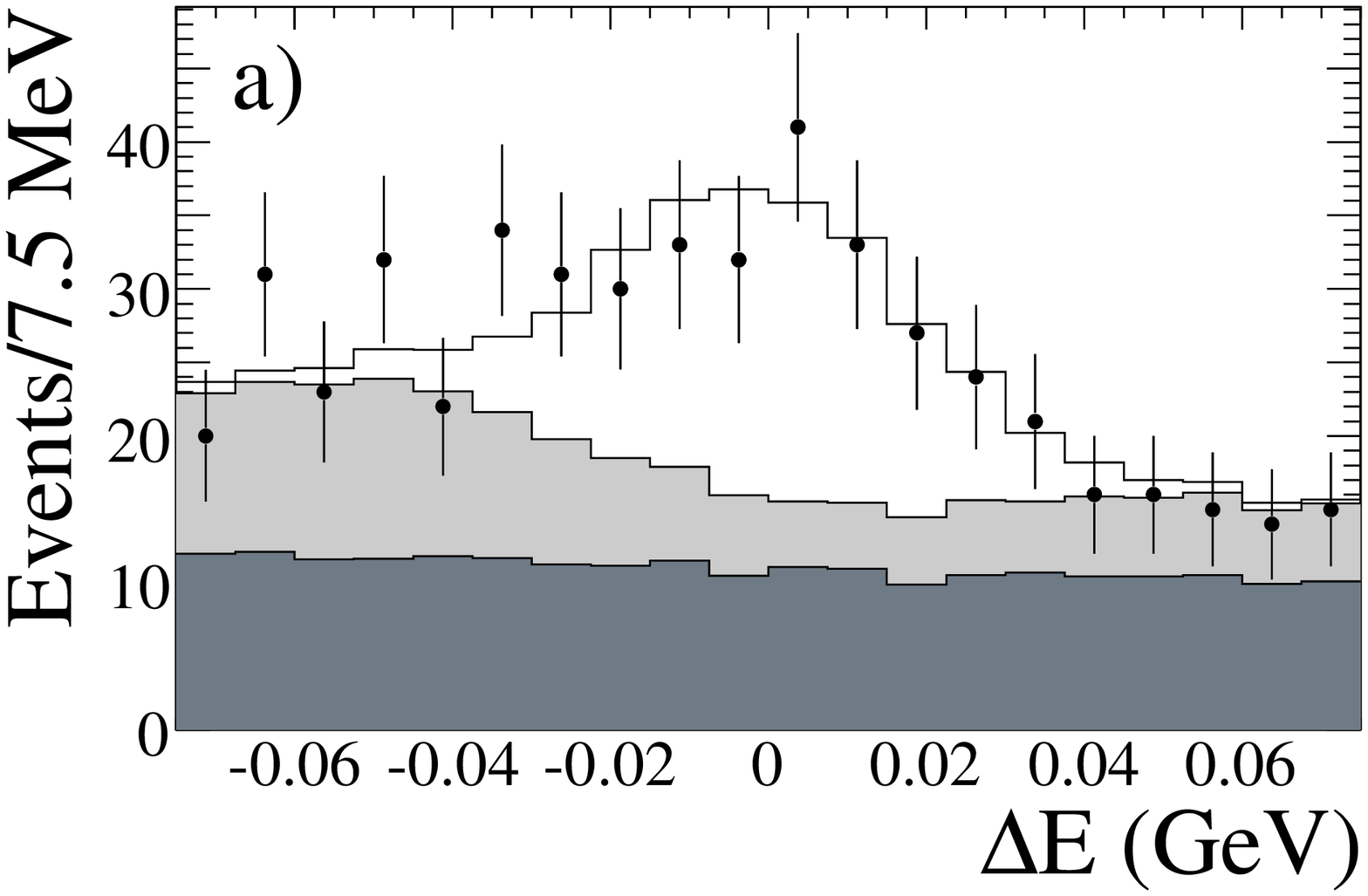}}
        {\includegraphics[width=.48\textwidth]{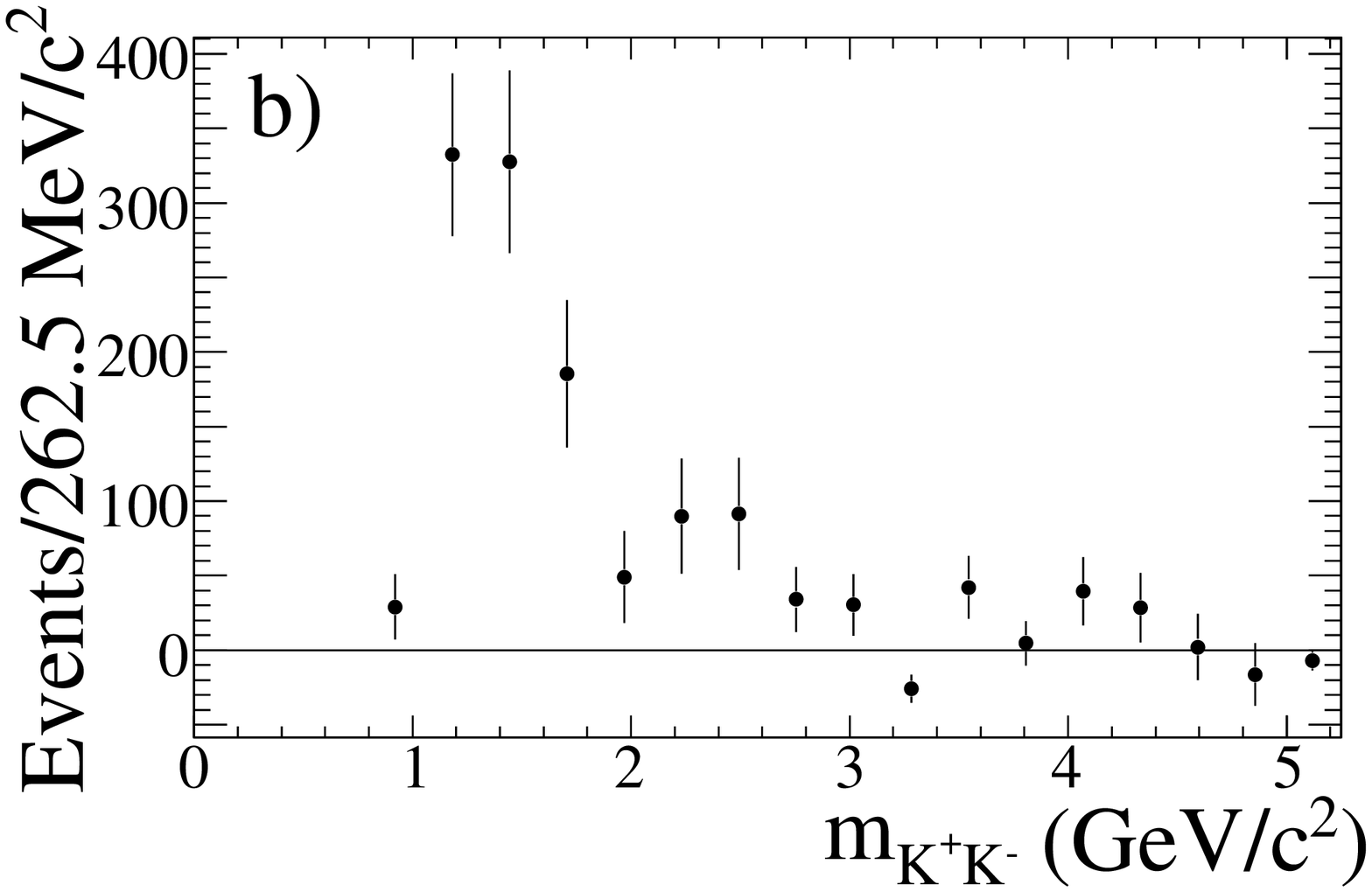}}
        \caption{\it a) \DeltaE\ projection of $B^+\to K^+K^-\pip$
        candidate events and b) Efficiency-corrected $m_{K^+K^-}$
        distribution of the $K^+K^-\pip$ signal candidates with
        $m_{K^+\pim}>1.5\gevcc$.}
\label{KKpifig}
    \end{center}
\end{figure}
\noindent Using 383 million \BB\ pairs recorded by \babar, we report the first
observation of charmless hadronic decays of charged \B\ mesons to the final state
$\Kp\Km\pip$~\cite{KKpi-paper}. We observe in total $429\pm43$ signal events with a
significance of 9.6 standard deviation ($\sigma$), and measure the inclusive branching
fraction~\cite{sPlot} ${\cal B}(B^+\to\Kp\Km\pip)=[5.0\pm0.5\stat\pm0.5\syst]\times10^{-6}$.
Figure~\ref{KKpifig}(a) shows the \DeltaE\ distribution of selected candidate events,
following a signal-enhancing requirement on the likelihood ratio which is formed
out of \mes\ and NN variables. Points show the data, the dark filled histogram shows
the \qqbar\ background and the light filled histogram shows the \BB\ background.
Approximately half of the signal events appear to originate from a broad structure
peaking near 1.5\,\gevcc\ in the $\Kp\Km$ invariant mass distribution (see
Figure~\ref{KKpifig}(b)). This structure is reminiscent of similar states
observed in Dalitz plot analyses of $B^+\to\Kp\Km\Kp$~\cite{KKK-paper} and
$B^0\to\KS\Kp\Km$~\cite{KKKs-paper}, and is likely to be of great interest for
the understanding of low energy hadronic bound states~\cite{Klempt}. Results on
the $\Km\pip$ mass spectrum are in reasonable agreement with a dedicated Q2B
analysis~\cite{KstarK-paper} that has put the most stringent upper limit (UL)
on $B^+\to\Kstarzb(892)\Kp$ and a first UL on $B^+\to\Kstarzb_0(1430)\Kp$ at
$1.1\times10^{-6}$ and $2.2\times10^{-6}$, respectively (all quoted ULs are
computed at 90\,\% confidence level). The measured charge asymmetry is found
to be consistent with zero.
\subsection{Vector-Vector Decay $B^0\to\Kstarz\Kstarzb$
}
\begin{figure}[!htb]
    \begin{center}
        {\includegraphics[width=.48\textwidth]{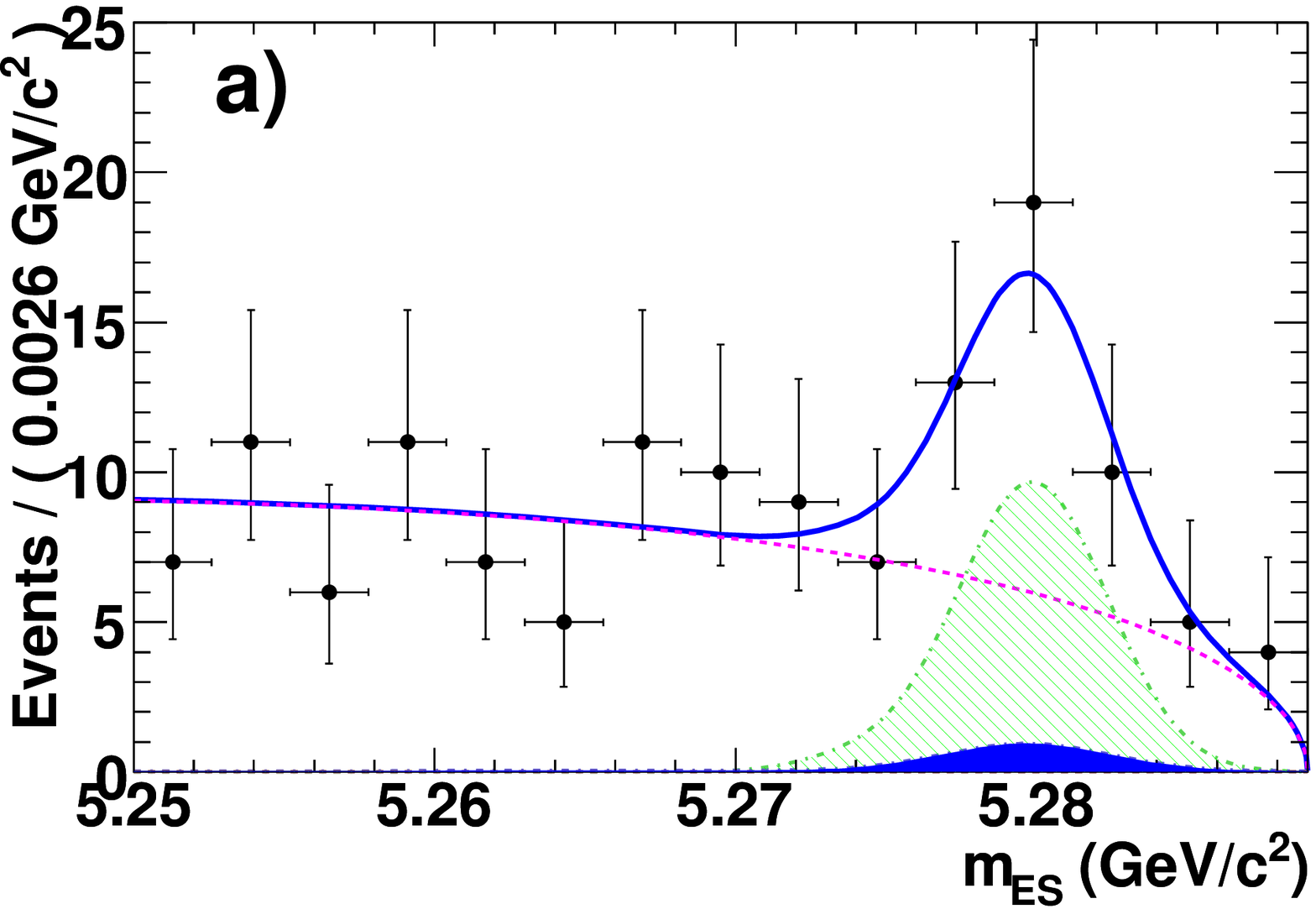}}
        {\includegraphics[width=.48\textwidth]{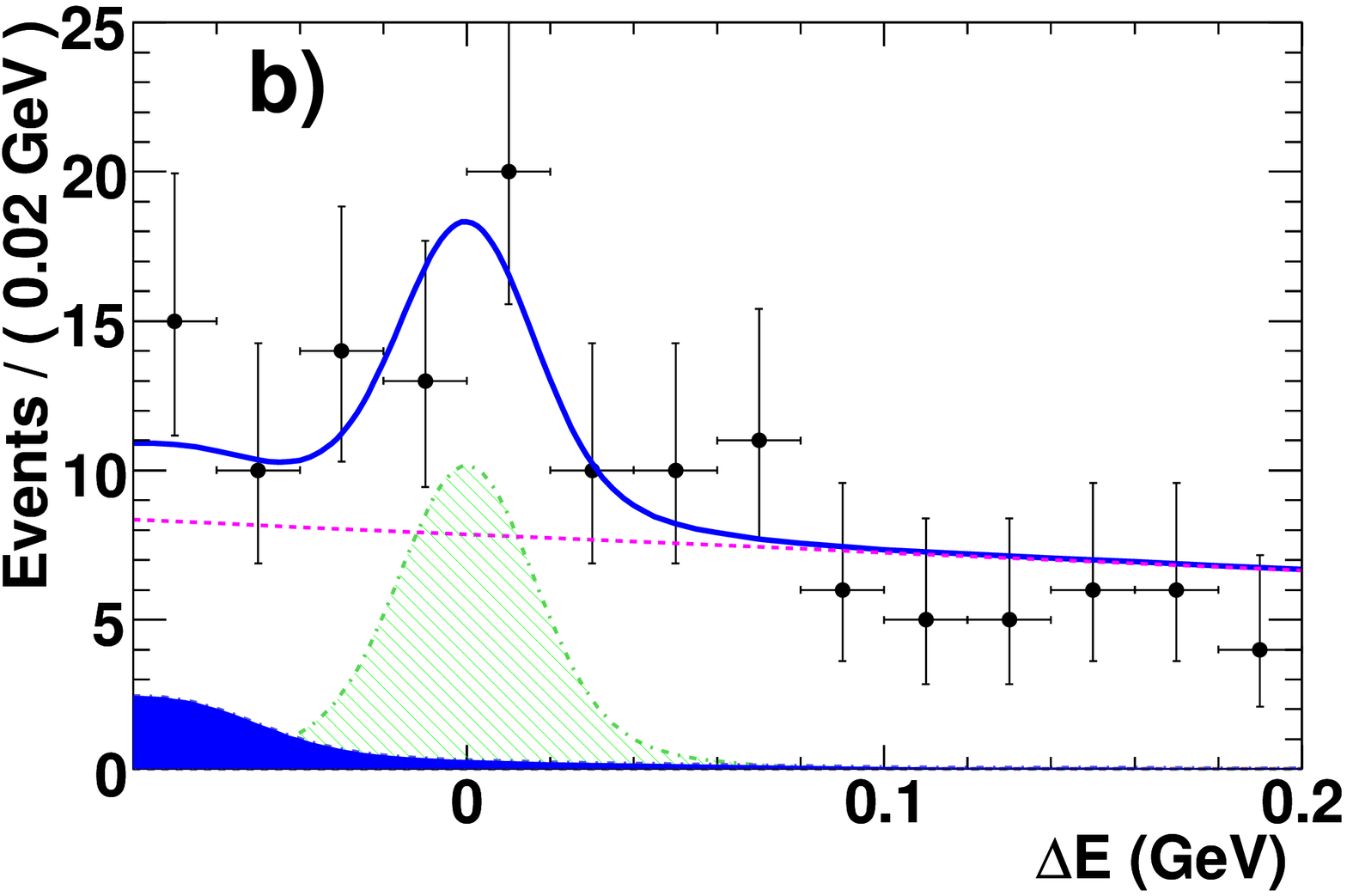}}
        \caption{\it Projections of a) \mes\ and b) \DeltaE\ of candidate
        events in the $B^0\to\Kstarz\Kstarzb$ decays.}
\label{KstKstbarfig}
    \end{center}
\end{figure}
\noindent We report the observation of the $\b\to\d$ penguin-dominated vector-vector
decay $B^0\to\Kstarz\Kstarzb$ at 6\,$\sigma$ significance~\cite{KstKst-paper}
with a sample of 383 million \BB\ pairs. Figure~\ref{KstKstbarfig} shows
\mes\ and \DeltaE\ projections of candidate events with a requirement on the
signal-to-background likelihood ratio, calculated excluding the plotted variable.
Points with error bars show the data, the solid line shows the projection for
signal-plus-background, the dashed line is the continuum background, the hatched
region is the signal, and the shaded region is the \BB\ background. Performing
a simultaneous likelihood fit we determine the branching fraction ${\cal B}
(B^0\to\Kstarz\Kstarzb)=[1.28^{+0.35}_{-0.30}\stat\pm0.11\syst]\times10^{-6}$, and
the fraction of longitudinal polarization $f_L=0.80^{+0.10}_{-0.12}\stat\pm0.06\syst$.
The branching fraction measurement is consistent with theoretical
predictions~\cite{KstKst-theo}. The measured $f_L$ value agrees with
the model expectation of QCD factorization~\cite{QCDF}, which predicts it
to be $\sim0.9$ in the vector-vector decay of a pseudoscalar ($B^0$ here).
We have also improved the existing UL on the SM-suppressed decay
$B^0\to\Kstarz\Kstarz$ by two orders of magnitude to $4.1\times10^{-7}$.
\subsection{Decays involving a Vector and two Pseudoscalars
}
Using a data sample of 383 million $\FourS\to\BB$ events, we measure the branching
fractions and \CP\ violation asymmetries~\cite{Ksthh-paper} of hadronic decays
$B^0\to\Kstarz h^+h^{\prime-}$, where $h$ and $h^\prime$ refers to either a kaon or
a pion. The $\Kstarz$ is detected via its self-tagging decay mode $\Kstarz\to\Kp\pim$.
Table~\ref{Ksthhtab} summarizes results of these measurements. We have made three
new observations in the final states $\Kstarz\Kp\Km$, $\Kstarz\pip\pim$ and
$\Kstarz\pip\Km$. In the SM-suppressed decay $\B^0\to\Kstarz\Kp\pim$, where
a branching fraction comparable to or larger than that of the $\Kstarz\pip\Km$
mode would be a signature of NP, no signal is observed and a first UL is set at
$2.2\times10^{-6}$. We find no evidence for \CP\ violation in these
$B^0\to\Kstarz h^+h^{\prime-}$ decays.
\begin{table}[!htb]
\centering
\caption{\it Measured signal yields, branching fractions (${\cal B}$),
significance ($S$) and \CP\ violation asymmetries (for the significant modes)
of $B^0\to\Kstarz(892)h^+h^{\prime-}$. The first uncertainty is statistical and
the second is systematic.
}
\vskip 0.1 in
\begin{tabular}{lcccc} \hline\hline
\multicolumn{1}{c}{$B^0\to$ Mode} & Signal yield & ${\cal B}(\times10^{-6})$ & $S(\sigma)$ & $A_{\CP}$ \\
\hline
$\Kstarz\KpKm$    & $984\pm46$ & $27.5\pm1.3\pm2.2$ & $>10$ & $0.01\pm0.05\pm0.02$\\
$\Kstarz\pip\Km$  & $183\pm42.4$ & $4.6\pm1.1\pm0.8$ & $5.3$ & $0.22\pm0.33\pm0.20$\\
$\Kstarz\Kp\pim$  & $18.8\pm29.4$ & $<2.2$ & $0.9$ & $-$\\
$\Kstarz\pip\pim$ & $2019\pm108$ & $54.5\pm2.9\pm4.3$ & $>10$ & $0.07\pm0.04\pm0.03$\\
\hline\hline
\end{tabular}
\label{Ksthhtab}
\end{table}
\subsection{Axial-vector and Pseudoscalar Modes
}
We report results of a first search for \B\ meson decays to final states
with an axial-vector meson, $b_1(1235)$, and a pseudoscalar meson (pion or
kaon), carried out with a data sample containing 382 million \BB\ pairs.
In the quark model the $b_1$ is the $I^G=1^+$ member of the $J^{PC}=1^{+-}$,
$^1P_1$ nonet. Its mass and width are ($1229.5\pm3.2$)\,\mevcc\ and
($142\pm9$)\,\mev, respectively, and the dominant decay is to
$\omega(782)\pi$~\cite{PDG2006}. By performing an extended ML fit to \DeltaE,
\mes, ${\cal F}$, and reconstructed invariant masses of the $b_1$ and $\omega$
resonances~\cite{b1h-paper}, we find clear signals for $B^+\to b^0_1\pip$,
$B^+\to b^0_1\Kp$, $B^0\to b^{\mp}_1\pi^{\pm}$ and $B^0\to b^-_1\Kp$.
Table~\ref{b1htab} summarizes the signal yield, branching fraction,
significance and charge asymmetry of these four decay modes. Measured
branching fractions and charge asymmetries agree with QCD factorization
predictions~\cite{b1h-theory}. Observations in the $B\to b_1K$ modes,
if confirmed with higher precision, would indicate a sizable weak
annihilation contribution to these modes~\cite{b1h-theory}. Furthermore,
the measurement of an asymmetry parameter,
$\Gamma(B^0\to b^+_1\pim)/\Gamma(B^0\to b^-_1\pip)=-0.01\pm0.12$ in the
decays $B^0\to b^{\mp}_1\pi^{\pm}$ agrees well with $G$-parity
suppression~\cite{Gparity}.
\begin{table}[!htb]
\centering
\caption{\it Measured signal yields, branching fractions (${\cal B}$),
significance ($S$) and direct \CP\ asymmetries of $B\to b_1(1235)h$ where $h=K/\pi$.
The first uncertainty is statistical and the second is systematic.}
\vskip 0.1 in
\begin{tabular}{lcrcr} \hline\hline
\multicolumn{1}{c}{Decay mode} & Signal yield &
\multicolumn{1}{c}{${\cal B}(\times10^{-6})$} & $S(\sigma)$ &
\multicolumn{1}{c}{$A_{\CP}$} \\
\hline
$B^+\to b^0_1\pip$ & $178^{+39}_{-37}$ & $6.7\pm1.7\pm1.0$ & $4.0$ & $0.05\pm0.16\pm0.02$\\
$B^+\to b^0_1\Kp$  & $219^{+38}_{-36}$ & $9.1\pm1.7\pm1.0$ & $5.3$ & $-0.46\pm0.20\pm0.02$\\
$B^0\to b^{\mp}_1\pi^{\pm}$ & $387^{+41}_{-39}$ & $10.9\pm1.2\pm0.9$& $8.9$ & $-0.05\pm0.10\pm0.02$\\
$B^0\to b^-_1\Kp$  & $267^{+33}_{-32}$ & $7.4\pm1.0\pm1.0$ & $6.1$ & $-0.07\pm0.12\pm0.02$\\
\hline\hline
\end{tabular}
\label{b1htab}
\end{table}

Moving on to another axial-vector state $a_1(1260)$, which is the $I^G=1^-$
state of the $J^{PC}=1^{++}$, $^3P_1$ nonet, we report evidence of two decay
modes containing pions in the final state~\cite{a1pi-paper} and two observations
in kaon modes~\cite{a1K-paper}. For pion modes the analysis comprises a smaller
dataset containing 232 million \BB\ pairs, while in kaon modes we have utilized
383 million $\FourS\to\BB$ events. Here the $a_1$ meson is reconstructed via
its most dominant decay to three pions. Neglecting contributions from
isoscalars, such as the $\sigma$ meson, to the two-pion state; we assume
${\cal B}(a^{\pm}_1(1260)\to\pi^{\pm}\pip\pim)$ is equal to ${\cal B}(a^{\pm}_1(1260)
\to\pi^{\pm}\piz\piz)$ and ${\cal B}(a^{\pm}_1(1260)\to(3\pi)^{\pm})=100\%$.
The three-pion decay is also considered as the only possible decay mode for
neutral $a_1$ mesons. These assumptions help in translating the product of
${\cal B}(B\to a_1(1260)h)$ and ${\cal B}(a_1(1260)\to3\pi)$ into the former.
Table~\ref{a1htab} summarizes this branching fraction measurement in the decays
$B^+\to a^+_1\piz$, $B^+\to a^0_1\pip$, $B^+\to a^+_1K^0$ and $B^0\to a^-_1\Kp$
along with the assorted signal yield and significance. Measured branching fractions
are in reasonable agreement with factorization model predictions~\cite{a1h-theory}.
In the case of kaon modes, we find no evidence for direct \CP\ violation.
\begin{table}[!htb]
\centering
\caption{\it Measured signal yields, branching fractions (${\cal B}$) and
significance ($S$) of $B\to a_1(1260)h$ where $h=K/\pi$. The first uncertainty
is statistical and the second is systematic.}
\vskip 0.1 in
\begin{tabular}{lcrc} \hline\hline
\multicolumn{1}{c}{Decay mode} & Signal yield &
\multicolumn{1}{c}{${\cal B}(\times10^{-6})$} & $S(\sigma)$ \\
\hline
$B^+\to a^+_1\piz$ & $459\pm78$ & $26.4\pm5.4\pm4.1$ & $4.2$ \\
$B^+\to a^0_1\pip$ & $382\pm79$ & $20.4\pm4.7\pm3.4$ & $3.8$ \\
$B^+\to a^+_1K^0$  & $241\pm32$ & $34.9\pm5.0\pm4.4$ & $6.2$ \\
$B^0\to a^-_1\Kp$  & $272\pm44$ & $16.3\pm2.9\pm2.3$ & $5.1$ \\
\hline\hline
\end{tabular}
\label{a1htab}
\end{table}
\section{Conclusions
}
\babar\ is pioneering several new measurements in charmless hadronic \B\ decays
that probe the SM in two orthogonal directions - the weak interaction by measuring
the CKM angles~\cite{sandrine}, and the strong interaction by exploring low-lying
hadronic bound states and by providing precision tests of QCD models. We eagerly
look forward to the last run, which together with data taken during the year
2006-2007 and not used in the presented results, would double the dataset. This
will be crucial for realizing other rare hadronic decay modes such as
$B^+\to\KS\KS\pip$ within our reach.
\section{Acknowledgments
}
It is a pleasure to thank the organizers of the HADRON07 conference for
a job well-done! This work is supported in part by the Science and Technology
Facilities Council of the United Kingdom, and the US Department of
Energy under contract number DE-AC02-76SF00515.

\begin{thebibliography}{99}
\bibitem{charge}Unless explicitly stated otherwise, charge-conjugate states are
assumed throughout.
\end{thebibliography}
\end{document}